\documentclass[aps,pre,eqsecnum,showpacs,draft]{revtex4}
\usepackage{amsmath,bm}
\begin{document}

\title{Structures and orientational transitions in thin smectic films of tilted hexatic}

\author{P.V. Dolganov}
\affiliation{Institute of Solid State Physics, RAS,  Moscow Region
 142432, Chernogolovka, Russia}
\author{K.I. Belov}
\affiliation{Institute of Solid State Physics, RAS,  Moscow Region
 142432, Chernogolovka, Russia}
\author{V.K. Dolganov}
\affiliation{Institute of Solid State Physics, RAS,  Moscow Region
 142432, Chernogolovka, Russia}
\author{E.I. Demikhov}
\affiliation{Institute of Solid State Physics, RAS,  Moscow Region
 142432, Chernogolovka, Russia}
\author{B.M. Bolotin}
\affiliation{Institute of Chemical Reactants and Especially Pure
Substances IREA, 107076 Moscow, Russia}
\author{E.I. Kats}
\affiliation {Laue-Langevin Institute, F-38042, Grenoble, France }
\affiliation {L.D. Landau Institute for Theoretical Physics, RAS,
117940 GSP-1, Moscow, Russia}

\date{\today}

\begin{abstract}
We present detailed systematic studies of structural
transformations in thin liquid crystal films with the smectic-$C$
(Sm$C$) to hexatic (HSm$B$) phase transition. For the first time
all possible structures reported in the literature are observed
for one material ($5 O.6$) at the variation of temperature and
thickness. In unusual modulated structures the equilibrium period
of stripes is twice with respect to the domain size. We interpret
these patterns in the frame work of phenomenological Landau type
theory, as equilibrium phenomena produced by a natural geometric
frustration in a system having spontaneous splay distortion.

\end{abstract}


\maketitle

\section{Introduction}
\label{int} Spontaneous formations of spatial patterns arise in a
wide variety of dynamic processes. Even more spectacularly they
are observed in equilibrium situations involving, fluids, solids
and liquid crystals. Especially remarkable are free standing
smectic films where one can avoid (in other systems often
dominating) influence of underlying substrates. These systems
provide a realization of many models describing diverse apparently
disparate physical phenomena (phase transitions, frustrations,
ferro-electricity, magnetism), and an opportunity to study the
crossover from two-dimensional to bulk behavior by drawing films
of increasing thickness.

For liquid crystal materials with phase sequences in bulk samples
HSm$B$ (or crystalline Cr$B$) -- Sm$A$ the phase transformations
in thin free-standing films are well studied at present time
\cite{SH95,JO03}. At temperature $T_{S1}$ which is about
${10^\circ C}$ above the bulk transition point $T_C$ the phase
transition occurs only in surface smectic layers. The interior
layers remain in the Sm$A$ phase. Below $T_{S1}$ there are no
phase transitions up to the temperature slightly above $T_C$. The
next transition in nearest to surface layer occurs at the
temperature $T_{S2}$ about $1^\circ$C above $T_C$. Sequence of
discrete layer-by-layer transitions on cooling may constitute from
2 to 5~transitions. In thick films the transition of the whole
film into the low temperature phase occurs at $T \sim T_C$.

Few words on nomenclature of tilted hexatic smectics
(usually labeled as Sm$F$,  Sm$I$, Sm$L$)
may be helpful here.
In the Sm$I$ phase the tilt direction is along of the local bonds,
in the Sm$F$ phase it is halfway between two local bonds, while in
Sm$L$ phases, the tilt direction lies along an intermediate
angle.
Transitions in the films of tilted smectics in which the high
temperature bulk phase is Sm$C$ and the low temperature phase is
tilted hexatic 
occur in an essentially different
way \cite{MS92,MS94,DD96,CP03}. 
In this case no sequence of
discrete layer-by-layer transitions is observed. After the surface
transition into the hexatic structure, the transformation of the film
structure on cooling continues in a broad temperature range.
Several characteristic temperatures $T_i$ may be pointed at which
qualitative changes of the film texture take place. Similar structure
and texture transformations were observed in several smectic
materials
\cite{MS92,MS94,DD96,CP03}.
So one may expect
that these transformations have common physical nature and occur
through a universal mechanism. However, up to now the mechanisms of
these structural transformations and physics behind remain unclear.

Our motivations for presenting this paper are twofold. First, in
section \ref{obs}, we present detailed systematic studies of
structural transformations in thin liquid crystal films with Sm$C$
to hexatic phase transition. We go one step further with respect
to the results known already (see, e.g.,
\cite{MS92,MS94,DD96,CP03}) investigating all possible structures
in one material ($5 O.6$) at the variation of temperature and
thickness. Besides, in section \ref{th} we rationalize and
interpret our observations in the frame work of simple
phenomenological model which includes the minimal number of
ingredients, i.e., it is just at the border between under-fitting
models (those that do not explain the data well) and over-fitting
models (those that fit the data too well by using too many
parameters). Although our model is a toy model in the sense of
caricaturizing some physical features, when properly interpreted,
it can yield quite reasonable values for a variety of measured
quantities. A more realistic model will not affect much our
conclusions, and transparency of treatment is worth a
simplification. The conclusion section is used to briefly
summarize our results and to augment their discussion.

\section{Observations}
\label{obs}

The measurements were made on the Schiff's-base compound
4-n-hexyl-N-[4-n-pentyloxy-benzilidene]-aniline ($5O.6$). The
sequence of phase transitions in the bulk sample is Sm$A$ --
($50.5^\circ$C) -- Sm$C$ -- ($49.5^\circ$C) - HSm$B$. Below the
HSm$B$ in the bulk sample the transition to a tilted hexatic
structure (Sm$F$) takes place. Free-standing films were prepared
by drawing the liquid crystal in a smectic phase across a circular
4mm hole in a thin glass plate. The experimental set-up enabled
simultaneous optical observations and reflectivity measurements.
The thickness of the film was determined by the reflected
intensity from the film in the "backward" geometry \cite{BW83}.
Observations of film structure and phase transitions were made
using polarized light reflected microscopy (PRLM) and depolarized
light reflected microscopy (DRLM) \cite{LN99}. The images were
recorded by a CCD camera. The orientational order parameter
$P_2=\frac{1}{2}\left(3\left<\cos^2\alpha\right> -1 \right)$
\cite{LMB83} was determined by optical absorption measurements. At
cooling $P_2$ changes from 0.75 to 0.8 in the Sm$A$ phase, is
about 0.82 in Sm$C$, and increases up to 0.92 in the hexatic
phase.

We performed investigations of thin smectic films starting from
thickness of 2~molecular layers. Figure~1 shows the temperatures
of the transitions in the films. Similar symbols denote the
temperatures of the transitions between the similar structures in
the films of different thickness. The high temperature part of the
phase diagram corresponds to the Sm$C$ structure. As it was
established early \cite{HP84} the temperature of the Sm$C$-Sm$A$
phase transition in free-standing films is essentially shifted to
higher temperature with respect to the bulk samples. Texture of
the film is characterized by a smooth spatial variation of the
${\bf {c}}$-director (Fig.~2a). The picture was taken in the part
of the film with a point topological defect, which is typical just
for Sm$C$ phase.

Upon cooling of the Sm$C$ film the first phase transition (filled
circles, transition into the Sm$T_1$ state, Fig.~1) leads to an
abrupt change of the film texture (Fig.~2b). In liquid crystalline
materials with the bulk hexatic phases the higher temperature
transition is associated with the phase transition of the surface
layers into the hexatic structure \cite{SH95,JO03}. In thick films
the high temperature shift of the transition with respect to the
phase transition temperature into the HSm$B$ structure in the bulk
sample is about 9$^\circ$C. This value is approximately the same
as for the HSm$B$--Sm$A$ transition \cite{SH95,JO03}. A different
situation is observed for thin films. While for the Sm$A$ phase
the shift of the transition in the 2-layer film with respect to
thick films does not exceed $2^\circ$C, this shift in our case is
essentially larger (more than $6^\circ$C). Below the transition
the film consists from domains with different ${\bf {c}}$-director
orientations and sharp boundaries between them (Fig.~2b). Such a
texture may be expected for tilted hexatic in which the ${\bf
{c}}$-director has a discrete set of orientations and
correspondingly sharp boundary between domains. However, it is not
typical for the Sm$C$ structure, which exists in the interior of
the film. For the films with $N>5$ the view of the films is mainly
determined by the Sm$C$ structure of the film interior. Thus, we
conclude that the Sm$C$ structure inside the film differs
sufficiently from the conventional Sm$C$ structure. Sharp boundary
between the domains suggests that not only in the surface layers
(i.e., in the hexatic state) but also inside the film (i.e., in
the Sm$C$ state) at the domain boundary the break of the ${\bf
{c}}$-director orientation occurs. The question arises about the
nature of the boundary between domains inside the film. These
peculiarities of the Sm$C$ structure inside the film become
extremely essential after the next transition (open squares in
Fig.~1).

At the next transition (into the Sm$T_2$ state, Fig.~1, Fig.~3)
the domains break up into narrow parallel stripes with alternating
brightness and sharp boundaries. According to
Refs.~\cite{MS92,MS94} at this transition the surface layers
transform into a so-called Sm$L$ phase in which the tilt plane is
oriented in the hexatic structure at the angle 15$^\circ$ with
respect to the direction of the hexatic bond orientation order. In
this structure there are possible 12 equivalent (i.e., having the
same energy) orientations of the tilt plane. Our optical
measurements confirm that the difference in the ${\bf
{c}}$-director orientation in the neighboring stripes is about
30$^\circ$ ($\pm$5$^\circ$) with two symmetric orientations of the
${\bf {c}}$-director relative to the stripe boundary $\varphi=\pm
15^\circ$ (Fig.~4a-c). The direction of the hexatic bond
orientational order does not change across the stripes (along
x-axis), while the direction of tilt plane changes at the stripe
boundary. Inside the film, in the Sm$C$ structure there is also a
break in the ${\bf {c}}$-director orientation. Contrary to this, a
smooth change of stripe orientation along domains is connected
with change of the direction of bond orientation order, meanwhile
the orientation of the ${\bf {c}}$-director relative to the bond
orientational order is preserved. The inset in Fig.~3a clarifies
the periodic stripe structure, in particular in the region of the
contact between two stripe domains. The stripe period increases
with decreasing temperature (Fig.~5). When upon cooling the stripe
period achieves the value of about 13--15$\mu$m, the stripe width
increases sharply and the structure becomes aperiodic. Open
triangles in Fig.~1 show the temperatures of this transition.

In the Sm$T_3$ region (Fig.~1) the film texture may be different
(Fig.~3b,c and Fig.~6a) and in many aspects is similar to the
observed earlier in films of Sm$C$ material with hexatic phases
\cite{MS92,MS94,DD96,CP03}. However the transition temperatures
between different textures are hardly reproducible. Moreover the
low temperature texture (Fig.~6a) as a rule transforms at heating
directly into the state with periodic stripes (Fig.~3a). Due to
this reason in the phase diagram (Fig.~1) we point only the
transition temperature between periodic stripe and aperiodic
structures (open triangles in Fig.~1). The domain structure, shown
in Fig.~3b, is formed on cooling from the narrow periodic stripes
through their broadening. This picture (Fig.~3b) was obtained by
means of depolarized light reflection microscopy. The domain
boundaries with the same brightness are oriented at about
45$^\circ$ with respect to the polarizers. It manifests that
orientations of the ${\bf {c}}$-director in the neighboring
domains are symmetric with respect to the domain boundary.
Measurements with crossed polarizes prove that the ${\bf
{c}}$-director in domains is oriented at the angles
$\pm$15$^\circ$ with respect to the domain boundary. Therefore,
the structure of wide domains (Fig.~3b) is similar to the narrow
periodic stripes (Fig.~3a). The honeycomb texture (Fig.~3c) forms
from the line domains (Fig.~3a) and exists only in a small
temperature range ($\leq 0.5^\circ$C). More typical textures in
the Sm$T_3$ state are domains with continuous change of the ${\bf
{c}}$-director orientation across the domains (upper part of
Fig.~6a) or large domains (lower part of Fig.~6a) also with a
continuous variation of the ${\bf {c}}$-director orientation. In
thin films this structure can be cooled to low temperatures ($<
40^\circ$C). In thick films ($N>10$) a reversible phase transition
is observed with formation of the texture shown in Fig.~6c. This
texture is typical for the tilted Sm$F$ phase. In the limit of
very thick films ($N>100$) crossed domains may be formed (Fig.~6b)
below the surface phase transition temperature. Two independent
sets of domains are formed at both film surfaces. Formation of
these independent structures in the thick films may point that
surface correlation length $\xi_S$ is less than about 50 smectic
interlayer periods.

\section{Theoretical interpretation}
\label{th} In order to provide a more complete account of the
phenomena described in the previous section, it would seem
appropriate to discuss how the observed results can be
consistently modeled theoretically. Without prior knowledge of
the actual structure, we shall assume the simplest model to answer
the natural questions why the phase transitions in materials with
Sm$A$ and Sm$C$ phases are so different, and what kind of
mechanisms are responsible for the formation of the periodic
stripe structure and its temperature dependence. From our
experimental observations a few conclusions about the following
qualitative features of the film structures and their
transformations seem inescapable.

First, formation of the periodic stripes (Fig.~3a) is related to
the structure of the surface layers. In the higher temperature
Sm$T_1$ state the ${\bf {c}}$-director is oriented along one
direction (Fig.~4a) in the middle between apexes of the Sm$F$
phase hexagon. In the Sm$L$ phase the energy minimum splits and
the ${\bf {c}}$-director may orient in two directions
\cite{MS92,MS94} corresponding to two equivalent energy minima
($\pm$15$^\circ$ with respect to the initial orientation,
Fig.~4a,b). As it is known competing attractive and repulsive
interactions generate domain patterns in a wide variety of systems
\cite{SA95}. Formation of a periodic structure is a signature of
instability that arises from a competition between two
antagonistic fields, and that, above some threshold, a modulated
state has a lower energy than the uniform one. In liquid crystals
this scenario is often related with existence of electric
polarization and a certain competition between the long-range
forces (namely electric and elastic ones). In the thin film under
consideration, the electric polarization may appear due to
nonuniform profile of the order parameter induced by the film
surface and also because the surface layers are in the hexatic
phase. This polarization $P_l$ is longitudinal (i.e., parallel to
the tilt plane) and points in opposite directions in the upper and
lower parts of the film \cite{AD98,AD99,LNM99}. These interactions
(electrostatic and elastic) contribute differently into the energy
of alternative configurations associated with existence of the
domain walls, and may lead to stabilization of the equilibrium
stripe period (as in solid crystal ferroelectric domains).

In liquid crystal films with the broken polar symmetry there is
also another cause of the stripe formation
\cite{MP73,BH85,LS86,HP89,SW93,KS01,VC05}. Indeed, the broken
polar symmetry allows linear over space gradients terms into the
Landau type free energy expansion. These terms affect the elastic
constants, which even may tend to zero. In this situation the free
energy of the defect structure may become more favorable than the
uniform structure. Thus the uniformly ordered state becomes
unstable with respect to the striped phase with periodic domain
walls. The equilibrium modulated structure arises to optimize the
gain in the elastic energy of the orientational deformation inside
the stripes and the energy cost to have the defect.

In literature devoted to theoretical descriptions of the modulated
phases in smectic films mostly models for polar smectic liquid
crystal films with transverse polarization are discussed
\cite{BH85,LS86,HP89,SW93,KS01}. Apparently it is not the case for
our system. There are several distinctions between the stripes
shown in Fig.~3a (see also their schematic representation in
Fig.~4b,c) and periodic stripes discussed in
\cite{BH85,LS86,HP89,SW93,KS01,VC05}. First, in our case the
structure of the surface layers is hexatic and it dictates the
value of director jump at the domain boundary. Second, the
neighboring stripes found in the works
\cite{BH85,LS86,HP89,SW93,KS01,VC05} have identical structures,
whereas in our case the azimuthal molecular orientation in the
adjacent stripes differs: ${\bf {c}}$-director is rotated
clockwise with respect to the symmetric orientation in the left
stripe (Fig.~4b,c) and counterclockwise in the right stripe
(Fig.~4b,c). Next, inside the stripes investigated in
\cite{BH85,LS86,HP89,SW93,KS01} elastic deformation is of the bend
type with the same sign of the bend in all stripes and with defect
walls in which the ${\bf {c}}$-director jumps back. In our case
the reorientation of the ${\bf {c}}$-director between stripes is
of the splay type. Moreover, there is no visible orientational
deformation of the ${\bf {c}}$-director inside stripes (Fig.~3a).
Thus care must be taken when comparing published theoretical
results to our experimental data. Below we examine one important
aspect of the liquid crystal modulated phase formation which does
not appear to have been investigated in any generality.

In our opinion, unusual structure of the stripes we have observed
is related to the nature of the geometrical frustration in the
films formed by non-chiral material. It was recognized quite some
time ago that because of the up-down asymmetry, the achiral
smectic film exhibits polar properties, in particular the ${\bf
{c}}$-director may be considered as a true vector (i.e., ${\bf c}$
and $-{\bf c}$ states are not equivalent). In ferroelectric
Sm$C^*$ phases the chiral asymmetry favors a bend
($\nearrow\longrightarrow\searrow$) in the ${\bf {c}}$-director
$\lambda_b \mathbf{\nabla}\times\mathbf{c}$. The preferred bend
direction (the sign of the coefficient $\lambda_b$
 at the linear over space derivatives term)
is determined by handedness of
the material (or by the direction of the ferroelectric polarization)
which is the same in the whole film.

We argue below, that in achiral systems the instability arises
from a competition between two elastic energies, the usual
quadratic Frank elastic energy, which favors a uniform orientation
of the ${\bf c}$-director, and an additional surface elastic term
linear in ${\bf c}$ gradient which promotes spontaneous splay
distortions. The linear over gradients terms like
$\mathbf{\nabla}\times\mathbf{c}$ are not allowed by the symmetry.
However, in such a film the broken chiral symmetry occurs as a
result of asymmetry between the surfaces (possessing hexatic
ordering) and the interior of the film (which is in the Sm$C$
state). Our measurements show that not only the tilt angle but
even the orientational order parameter $P_2$ differs essentially
in Sm$C$ and hexatic structures. The surface-induced term linear
over the gradients of \textbf{c} ($ \propto \lambda_s
\mathbf{\nabla}\cdot\mathbf{c}$) favors a splay deformation
($\nwarrow\uparrow\nearrow$) in the \textbf{c} field and has the
opposite signs (direction of splay curvature) in the top and
bottom parts of the film \cite{MS92,MP73}. The ${\bf c}$-director
is also the order parameter of the film. It allows one to describe
its macroscopic physics, in particular, to write its free energy
in the spirit of the Landau theory
\begin{eqnarray}
\label{k1} F = \frac{1}{2}K_s(\nabla\cdot {\bf {c}})^2 +
\frac{1}{2}K_b (\nabla \times {\bf {c}})^2 + \lambda _s
\nabla\cdot {\bf {c}} + \frac{1}{2} A {\bf {c}}^2 + \frac{1}{4}B
{\bf{c}}^4 \, .
\end{eqnarray}
The first two terms are the splay and bend elastic energies, and
Frank elastic moduli $K_s$ and $K_b$ are proportional to the film
thickness. The last two terms are conventional Landau expansion.
The third term $\nabla \cdot {\bf {c}}$ is a total derivative that
can be transformed to boundary terms. Therefore it is relevant
only for the thin films. For the thick films another term with the
same symmetry can be constructed
\begin{eqnarray}
\label{k2}
 \lambda _s^\prime {\bf {c}}^2  \nabla \cdot {\bf {c} }
\, ,
\end{eqnarray}
and to avoid its reduction to the pure surface  contribution,
variations in the ${\bf {c}}$ amplitude are needed. These kinds of
the contributions into the free energy (terms linear over the
splay distortion $ \mathbf{\nabla}\cdot\mathbf{c}$) lead to
formation of the unusual modulated structure we have observed in
this work.

If the molecules forming the system carried permanent dipole
moments {\mbox{\boldmath{$\mu $}}} with a non-vanishing component
along the direction ${\bf {c}}$, then the phase would exhibit
spontaneous electric polarization ${\bf {P}}$. This spontaneous
polarization is proportional to the polar order parameter. For
simplicity and for the lack of different compelling indications
from the experimental part of our work, the dipolar forces will be
neglected in the stripe period estimation below. 
It might be the case if the molecules involved have relatively
large shape anisotropy (and not a large electric dipole
moment), and ionic impurities screen the Coulomb interaction.
However, it is
not the whole story. In order to get the correct structure of the
splay phase, one has to take into account the complete order
parameter, including the modulus $|{\bf {c}}|$. Besides, there is
a price to be paid, because $|{\bf {c}}|$ can not be constant,
where the splay is constant. Indeed, in 2D the splay distortion of
the orientation can not occur in a defect-free fashion. Instead,
to relieve this frustration, the system will form a modulated
phase consisting of a regular network of defect walls and points.

It is interesting and tempting to hypothesize the following stripe
structure satisfying such kind of the symmetry breaking. The
uniform Sm$C$ structure of the top part of the film breaks up into
finite regions with splay deformation of the ${\bf {c}}$-director
(Fig.~4d). Regions with the same favorable sign of the splay
(counterclockwise in Fig.~4d) are separated by defect lines in
which the ${\bf {c}}$-director abruptly rotates back. In the
bottom part of the film (Fig.~4e) the ${\bf {c}}$-director
rotation is the opposite (clockwise). The lines in which the ${\bf
{c}}$-director jumps back are shifted in the x-direction on the
stripe period $d_{st}$ with respect to the top part of the film
(Fig.~4d,e). In this structure the direction of the splay
modulation in \textbf{c} is favorable on the both sides of the
film. Remarkably the net magnitude of the ${\bf {c}}$-director
orientation $\varphi$ across the film is constant in each stripe
in the agreement with our experimental data. Their values
$\varphi=\pm 15^\circ$ are dictated by structure of the surface
Sm$L$ phase. To be stable a splay modulated structure has to
overcome the unfavorable core defect energy ($\varepsilon $) and
the ordinary nematic order parameter contribution. In the zero
approximation the periodic stripe phase exists when the gain in
the surface-elasticity energy exceeds the energy $\varepsilon$
 of the domain wall.
Competition between these energies determines the
stripe width \cite{MP73}:
\begin{eqnarray}
d_{st} \approx \frac{K}{\lambda_s-\varepsilon} , \label{width}
\end{eqnarray}
where $K=(1/2)(K_s + K_b)$ is the mean Frank constant. In the
temperature window $T > T_C$ ($T_C$ is the bulk Sm$C$--HSm$B$
transition temperature) we are interested in, the main temperature
dependent factor in (\ref{width}) is $\lambda _s$. The very
existence of this linear splay distortion is due to the asymmetry
of the order parameter profile over the film. Induced by the
surface ordering $\Psi _s$, hexatic (bond) order parameter $\Psi $
decays toward the interior of the film
\begin{eqnarray}
\label{k3}
 \Psi (z) \propto \Psi _s \frac{\cosh [(z-L)/\xi ]}{\cosh (L/\xi )}
\, ,
\end{eqnarray}
where $L$ is the film thickness, and $\xi $ is the bulk phase transition correlation length
\begin{eqnarray}
\label{k4}
 \xi \simeq \frac{\xi _0}{\sqrt {(T-T_C)/T_C}}
\, ,
\end{eqnarray}
with $\xi _0$ designating the bare microscopic correlation length.
The asymmetry of the profile $\Psi $ determines the value of the parameter $\lambda _s$
\begin{eqnarray}
\label{k5}
 \lambda _s \propto \tanh (L/\xi )
\, ,
\end{eqnarray}
and the transition from the homogeneous to the modulated phase
occurs if the asymmetry is strong enough (see, (\ref{width})).
Thus we end up with the conclusion
that the stripe phase (soliton regime) appears
spontaneously on cooling and then the period of the distortion
increases with decreasing temperature.
The second conclusion from (\ref{width}) is that upon cooling
when $\lambda _s$ exceeds $\varepsilon $, the periodic stripe structure
becomes unstable.
The both conclusions conform to our experimental
observations.

We observe complex phase behavior with various equilibrium structures.
A separate question how to calculate all equilibrium structures of a specific system,
requires full minimization of the global free energy, and it depends on unknown
phenomenological Landau expansion coefficients. We will not even attempt the
calculation of such a complex phase diagram in this paper but will content
ourselves with one remark.
Modulated phases we found have nonuniform density or orientation
distributions, that is to say
their symmetry is that of a solid or a liquid crystal.
The difference between the phases which appear under the name of modulated structures, and solids
or liquid crystals, is that generally the period of modulation
is larger.

\section{Conclusion}
\label{con} 

It is not a major goal of this work to achieve quantitative agreement between the results
obtained with our phenomenological model and experimental measurements.
However, since the present understanding of the mechanism leading to modulated structures in achiral
tilted hexatic films is incomplete, the
model may be an appropriate tool for working out typical trends that may be testable in experiments.
In this paper we have presented results of studies of
structural transformations in thin liquid crystal films with the
smectic-$C$ (Sm$C$) to hexatic (HSm$B$) phase transition, and
their interpretation within a simple phenomenological model. The
free energy was written in the most simple form that involves the
least number of model parameters, and we have shown that this
simple model can capture many of the features seen in experiment.
Our interpretation of the results is based on the simple
consideration that because of the up-down asymmetry, the achiral
smectic film exhibits polar properties. One note of caution is in
order here. In fact the structure of any Sm$C$ phase is inherently
polar since the tilt singles out a unique direction about the
layer normal {\mbox{\boldmath{$\nu $}}} (although, the directors ${\bf {n}}$
and $-{\bf {n}}$ are physically indistinguishable). Therefore in
the Sm$C$ structure one may have a pseudo-vector
\begin{eqnarray}
\label{k6}
{\bf {l}} = ({\mbox{\boldmath{$\nu $}}} \times {\bf {n}})({\mbox{\boldmath{$\nu $}}}{\bf {n}})
\, .
\end{eqnarray}
Obviously ${\bf {l}}$ is perpendicular to the tilt plane, and this kind
of the polarity (along the pseudo-vector ${\bf {l}}$) is of fundamentally
different nature from the polarity along ${\bf {c}}$ we investigated in our paper.
Indeed, ${\bf {l}}$ polarity is compatible with mirror
symmetry in the tilt plane, whereas the polar splay distortion responsible
for the stripe
structure, changes its sign under such a mirror reflection.

A question of primary importance is the understanding of the
origin of the thermodynamic behavior we found in our work. It is
well known \cite{BI86} that for a film with the uniform ordering
(like nematic or ferromagnetic), when the interaction at the
boundaries is such that it enhances local order, it may happen
that a surface transition takes place at temperatures above the
critical temperature of the bulk. In such a transition, the layers
close to the surface become ordered although the bulk remains
disordered. Depending on the nature of the interactions between
the bulk and the surface, the system may exhibit various surface
phase transitions, for instance, wetting transitions. In the
latter case at temperatures just below the bulk transition, the
thickness of the surface ordered layer is infinite. Unlike this
scenario, non-uniformly ordered (modulated) systems not
necessarily exhibit wetting phenomena, in which the thickness of
surface ordered layer diverges. Instead of it, the system might
exhibit a transition from one surface state to another, where both
surface states have a finite thickness \cite{SA95}. Since we have
to deal in this work with modulated (non-uniform) structures,
aforesaid arguments provide the physically appealing thermodynamic
interpretation of our results.

First of all since we have deal with films, the basic
thermodynamics of phase transitions should be formulated for this
kind of the restricted geometry. Such a problem has been discussed
for various systems long ago, and the results borrowed from the
textbooks \cite{BI86,LL80} read as follows. In a film of thickness
$L$, the thermodynamic potential $G$ per unit area is
\begin{eqnarray}
\label{d1} \frac{G}{A} = - p L + 2 \gamma \, ,
\end{eqnarray}
where $\gamma $ is the surface free energy, $p$ is the bulk
pressure, and $A$ is the area. We denote the surface free energies
of the smectic $C$ and hexatic phases by $\gamma _C$, and $\gamma
_H$ respectively, and the Laplace condition yields
\begin{eqnarray}
\label{d2} \gamma _C = \gamma _H + \gamma _{C-H}\cos \theta \, ,
\end{eqnarray}
where $\theta $ is the contact angle between Sm$C$ and hexatic,
and $\gamma _{C - H}$ is the surface energy at the interface. In
the bulk the coexistence temperature $T_C$ is defined by the
equilibrium condition
\begin{eqnarray}
\label{d3} p_C(T_C) = p_H(T_C) \, .
\end{eqnarray}
In the finite thickness $L$ film
\begin{eqnarray}
\label{d4} T_m = T_C + \Delta T(L) \, ,
\end{eqnarray}
and from $G_C = G_H$ one finds
\begin{eqnarray}
\label{d5} -p_H(T_m) + \frac{2\gamma _H}{L} = - p_C(T_m) +
\frac{2\gamma _C}{L} \, .
\end{eqnarray}
Since
\begin{eqnarray}
\label{d6} p = p(T_C) + S(T_C)\Delta T
\end{eqnarray}
($S$ is the entropy) we get
\begin{eqnarray}
\label{d7} \Delta T(L) = \frac{2(\gamma _C - \gamma _H)T_C}{L Q}
\, ,
\end{eqnarray}
and $Q = T_C(S_C - S_H)$ is the latent heat at the bulk
transition. The same can be expressed (see (\ref{d2})) as
\begin{eqnarray}
\label{d8} \Delta T(L) = \frac{2 T_C \gamma _{C-H}\cos \theta }{L
Q} \, .
\end{eqnarray}
However, in (\ref{d7}) we did not consider the interaction $\Pi
(L)$ \cite{POK01} between the film surfaces (or between the
walls). Repeating the same thermodynamics as above, one ends up
with the same kind of the equation but where the disjoining
pressure $p_d$ \cite{POK01} renormalizes the surface energy
$\gamma $
\begin{eqnarray}
\label{d11} 2\tilde \gamma (L) = 2 \gamma + L p_d + \Pi \, .
\end{eqnarray}
According to (\ref{d8}) when the contact angle $\theta =0$, the
HSm$B$ phase is favored near the boundary, when $\theta = \pi $,
the Sm$C$ phase is favored, and the intermediate values of $\theta
$ apply to intermediate structures. In any case for the finite
film thickness there are two possible scenarios depending on the
film thickness. In one scenario the phase transition will occur
before the surface HSm$B$ layer has had a chance to grow thick. In
the another scenario, one can have a phase transition at a
temperature at which HSm$B$ thickness at the surface is already
larger than the sample thickness $L$.

There are clearly several open questions and future challenges.
One of them is related to dipolar forces. Indeed, since the
molecules are tilted, and the interface and the interior symmetry
of the film are different from each other, the film has only one
symmetry element, the vertical mirror plane, which is
perpendicular to the film and parallel to the molecules. The film
is therefore equivalent to a two-dimensional polar nematic, and it
bears an electric polarization. If this polarization is small
compared to the elastic energies involved, our arguments given
above apply. On the other hand, if electric energies dominate,
another kind of the texture can be stable, since all splay centers
act as charge centers, and in this case a lattice of several small
charges is more stable than one big charge. Another interesting
question is how to tune parameters of the various modulated
structures to optimize properties of technological interest. For
instance, a wide area of research is clearly the problem to what
extent the investigated systems can be useful to achieve
interesting electro-optical properties. The treatment above can be
generalized to more realistic systems (e.g. dipolar forces
including), with the same conceptual ingredients, albeit at the
expense of a rapidly increasing complexity.

This work has been supported in part by RFFI grant No.~05-02-16675
and Program of Presidium RAS "Influence of atomic and crystalline
structure on properties of condensed media".

\newpage

\newpage

{\bf Figure Captions}

Fig.~1 \\ Temperatures of transitions observed in $5O.6$ films of
different thickness. The high-temperature state corresponds to the
Sm$C$ phase. Closed symbols denote structural transitions, open
symbols transitions associated with change of director orientation
in the film. In thick films ($N=17$) an additional transition to
Sm$T_4$ state takes place (filled diamond) with texture typical
for the Sm$F$ phase.

Fig.~2 \\ High temperature textures in a 7-layer film: Sm$C$,
$T=60.1^\circ$C (a), Sm$T_1$, $T=59.2^\circ$C  (b). In frame (a) a
point topological defect with characteristic brushes is observed.
DRLM. The horizontal size of the frames is about 420$\mu$m.

Fig.~3 \\ State with narrow periodic stripes, $T=55^\circ$C, $N=7$
(a). Inset: stripes in the same film after cooling to
$T=52.3^\circ$C. Structure with linear aperiodic domains,
$T=51.3^\circ$C, $N=17$ (b). Honeycomb texture may form on cooling
in a narrow temperature range (c), $T=50.8^\circ$C , $N=7$. The
horizontal size of the frames (a) and (b) is about 720$\mu$m,
frame (c) 480$\mu$m, and inset 160$\mu$m, DRLM.

Fig.~4 \\ Schematic representation of stripe structure. Stripes
are oriented along the y-axis. (a) Monodomain state (Sm$F$). (b)
In the Sm$L$ phase the $\bf {c}$-director may have two
orientations with respect to bond orientational order
($\varphi=\pm 15^\circ$). (c) Net orientation of the $\bf
{c}$-director in the stripes with jump in director orientation on
the stripe boundary (Sm$T_2$ state, Fig.~3a). Structure of the
Sm$C$ top (d) and bottom (e) layers of the film in the stripe
state.

Fig.~5 \\ Temperature dependence of the stripe period in a
four-layer film. The period monotonically increases with
decreasing temperature. The solid line is a fit to the eye.

Fig.~6 \\ Structures formed in films at low temperatures.
Aperiodic domains, $N=7$, $T=50.8^\circ$C (a). Two sets of
crossing domains in a thick film, $N=200$, $T=49.1^\circ$C (b). On
further cooling a transition to the Sm$F$ structure occurs (c)
$N=17$, $T=48.1^\circ$C, DRLM. The horizontal size of the images
is $374\mu$m.

\end{document}